\documentstyle[prl,epsfig,aps]{revtex}
\input{psfig.sty}
\begin{document}
\pagenumbering{arabic}
\pagestyle{plain}
\title{
Trajectories in phase diagrams, growth processes and computational 
complexity: \\
how search algorithms solve the 3-Satisfiability problem.}

\author{Simona Cocco \footnote{Current address: Department of Physics,
The University of Illinois at Chicago, 845 W. Taylor St., Chicago
IL 60607.} 
and R{\'e}mi Monasson \footnote{Current address: The James Franck Institute,
The University of Chicago,
5640 S. Ellis Av., Chicago IL 60637.} }
 
\address{CNRS-Laboratoire de Physique Th{\'e}orique de l'ENS, 
24 rue Lhomond, 75005 Paris, France }
 
\maketitle
 
\begin{abstract}
\widetext
Most decision and optimization problems encountered in practice fall
into one of two categories with respect to any particular solving
method or algorithm: either the problem is solved quickly (easy) or
else demands an impractically long computational effort (hard).  
Recent investigations on model classes of problems
have shown that some global parameters, such as the ratio between the
constraints to be satisfied and the adjustable variables, are good
predictors of problem hardness and, moreover, have an effect analogous
to thermodynamical parameters, e.g. temperature, in predicting phases
in condensed matter physics [Monasson et al., Nature 400 (1999)
133-137]. Here we show that changes in the values of such parameters
can be tracked during a run of the algorithm defining a trajectory
through the parameter space. Focusing on 3-Satisfiability, a
recognized representative of hard problems, we analyze trajectories
generated by search algorithms using growth processes statistical
physics.  These trajectories can cross well defined phases,
corresponding to domains of easy or hard instances, and allow to
successfully predict the times of resolution.
\end{abstract}  

\vskip .5cm
PACS Numbers~: 05.10, 05.70, 89.80 
\vskip .5cm



Consider a set of $N$ Boolean variables  (that can be either true
or false) and a set of $M=\alpha\; N$ constraints (called clauses),
each of which being the logical OR of three variables or of their
negations, see Figure~1. Then, try to figure out whether there
exists or not an assignment of variables satisfying all clauses
(called solution).

This problem, called 3-Satisfiability (3-SAT), is among the most
difficult ones to solve as its size $N$ becomes large.  It is a
fundamental conjecture of computer science that no method exists to
solve 3-SAT efficiently\cite{NPC}, {\em i.e.}  in time growing at most
polynomially with $N$. In practice, one therefore resorts to methods
that need, {\em a priori}, exponentially large computational
resources. One of these algorithms, the ubiquitous
Davis--Putnam--Loveland--Logemann (DPLL) solving
procedure\cite{DP,Cra,Hay}, is illustrated in Figure~1. DPLL operates
by trials and errors, the sequence of which can be graphically
represented as a search tree made of nodes connected through edges,
see Figure~1. Examples of search trees for satisfiable (sat) or
unsatisfiable (unsat) instances are shown Figure~2. Computational
complexity is the amount of operations performed by the solving
algorithm; we follow the convention that it is measured by the size of
the search tree, {\em i.e.} the number of nodes.


Complexity may, in practice, vary enormously with the instance, that is, the 
set of clauses,  under
consideration.  To understand why instances are easy or hard to solve,
computer scientists have focused on model classes of 3-SAT instances.
Probabilistic models, that define distributions of random instances
controlled by few parameters, are particularly useful to shed light on
the onset of complexity.  An example, that has attracted a lot of
attention over the past years, is random 3-SAT: all clauses are drawn
randomly and each variable negated or left unchanged with equal
probabilities.  Experiments\cite{Cra,AI} and theory\cite{MZ} indicate
that clauses can almost surely always (respectively never) be
simultaneously satisfied if $\alpha$ is smaller (resp. larger) than a
critical threshold $\alpha _C \simeq 4.3$ as soon as $M,N$ go to
infinity at fixed ratio $\alpha$. This phase transition\cite{Sta,MZ}
is accompanied by a drastic peak of hardness at
threshold\cite{Cra,AI}, see Figure~3. A complete understanding of 
this pattern of complexity has been lacking so far.

In this letter we argue that search algorithms induce a dynamical
evolution of the computational problem to be solved.  We shall show
how this complex and non-Markovian dynamics can be closely related to
(surface) growth processes. Statistical mechanics concepts and tools
thus allow to predict analytically and understand computational
complexity. For the sake of clarity, we shall report technical details
in another and extended paper \cite{ric}.


As shown in Figure~1, the action of DPLL on an instance of 3-SAT
causes the reduction of 3-clauses to 2-clauses. We use a mixed 2+p-SAT
distribution\cite{Sta}, where $p$ is the fraction of 3-clauses, 
to model what remains of the input instance at a node of the
search tree. Using experiments and methods from statistical
mechanics\cite{MZ}, the threshold line $\alpha _C (p)$ may be obtained
with the results shown in Figure~4 (full line).  The phase diagram of
2+p-SAT is the natural space in which the DPLL dynamic takes place. An
input 3-SAT instance with ratio $\alpha$ shows up on the right
vertical boundary of Figure~4 as a point of coordinates $(p=1,\alpha
)$.  Under the action of DPLL, the representative point moves aside
from the 3-SAT axis and follows a trajectory. The
location of this trajectory in the phase diagram allows a precise
understanding of the search tree structure and of complexity.
(in the following, we consider the trajectories generated by DPLL using
the GUC heuristics\cite{Fra} -- see caption of Figure~1 -- but our 
calculation could be repeated for other rules\cite{Cra,AI,Fra,Fri}). 


At sufficiently small ratios $\alpha < \alpha _L \simeq 3.003$, DPLL
easily finds a solution\cite{Fra,Fri}.  The search tree has a unique
branch (sequence of edges joining the top node to the extremity).
This branch grows as the fraction $t$ of variables assigned by DPLL
increases and terminates with a solution, see Figure~2A. Each node
along the branch carries a 2+p-SAT instance with characteristic
parameters $p$ and $\alpha$. The knowledge of the latter as functions
of the ``depth'' $t$ of the node, first established by Chao and
Franco\cite{Fra}, allows us to draw the trajectory followed by the
instance under the action of DPLL in Figure~4.  The trajectory,
indicated by a light dashed line, first heads to the left and then
reverses to the right until reaching a point on the 3-SAT axis at
small ratio without ever leaving the sat region. Further action of
DPLL leads to a rapid elimination of the remaining clauses and the
trajectory ends up at the right lower corner S, where a solution is
achieved. Thus, in the range of ratios $\alpha < \alpha _L \simeq
3.003$, 3-SAT is easy to solve: the computational complexity scales
linearly with the size $N$, see Figure~3.


For ratios above threshold ($\alpha > \alpha _C\simeq 4.3$), instances
have (almost) never a solution but a considerable amount of
backtracking is necessary before proving that clauses are
incompatible. As shown in Figure~2B, a generic unsat tree includes
many branches. The number $B$ of branches is related to the number $Q$
of nodes through the relation $Q=B-1$ valid for any complete tree,
check Figure~2B. Complexity grows exponentially\cite{Chv} as $2^{N
\omega}$. We have experimentally counted directly $Q$ or,
alternatively, $B$, and averaged the corresponding logarithms $\omega$
over a large number of instances. Results have then be extrapolated to
the $N\to \infty$ limit \cite{ric} and are reported in Table~1.

We have analytically computed $\omega$ as a function of $\alpha$,
extending to the unsat region the probabilistic analysis of DPLL. The
search tree of Figure~2B is the output of a sequential process: nodes
and edges are added by DPLL through successive descents and
backtrackings.  We have imagined a different building up, that results
in the same complete tree but can be mathematically analysed: the tree
grows in parallel, layer after layer. A new layer is added by
assigning, according to DPLL heuristic, one more variable along each
living branch. As a result, some branches split, others keep growing
and the remaining ones carry contradictions and die out. At a given
depth (fraction of assigned variables) $t$ in the tree, each branch
carries a 2+p-SAT instance, with characteristic parameters
$p$,$\alpha$. Its further evolution can be well approximated by a
Markovian stochastic process that depends only on $p$ and
$\alpha$. This approximation permits to write an evolution equation
for the logarithm $\omega(p,\alpha;t)$ of the average number of
branches with parameters $p,\alpha$ as the depth $t$ increases,
\begin{equation}
\frac{\partial \omega } {\partial t} = {\cal H} \left[ p, \alpha,
\frac{\partial \omega } {\partial p} , \frac{\partial \omega }
{\partial \alpha} ,t \right] \qquad , \label{croi}
\end{equation}
${\cal H}$ incorporates the details of the splitting
heuristics\cite{nota1}. Partial differential equation (\ref{croi}) is
analogous to growth processes encountered in statistical physics
\cite{Gro}.  The surface $\omega$, growing with ``time'' $t$ above the
bidimensional plane $p,\alpha$, describes the whole branches
distribution. Exponentially dominant branches are given by the top of
the distribution; the coordinates $p(t),\alpha(t)$ of the maximum of
$\omega$ define the tree trajectories on Figure~4. The hyperbolic line
indicates the halt points, where contradictions prevent tree
trajectories from further growing.  Along the trajectory, $\omega$
grows from 0, on the right vertical axis up to some final positive
value on the halt line.  This value is our theoretical prediction for
the logarithm of the complexity (divided by $N$).  Values of $\omega$,
obtained for $4.3<\alpha<20$ by solving equation (\ref{croi}) compare 
very well with numerical results, see Table~1.


The intermediate region $\alpha_L < \alpha<\alpha_C$ juxtaposes the
two previous behaviours, see tree Figure~2C. The branch trajectory,
started from the point $(p=1,\alpha)$ corresponding to the initial
3-SAT instance, hits the critical line $\alpha_c(p)$ 
at some point G with coordinates ($p_G,\alpha_G$).  The
algorithm then enters the unsat phase and generates 2+p-SAT instances
with no solution. A dense subtree, that DPLL has to go through
entirely, forms below G till the halt line, see tree trajectory in
Figure~4. From our theoretical framework, the logarithm $\omega$ of
the size of this subtree can be analytically predicted. G is the
highest backtracking node in the tree (see Figure~2C) reached back by
DPLL, since nodes above G are located in the sat phase and carry
2+p-SAT instances with solutions. We have checked experimentally this
scenario for $\alpha =3.5$. The coordinates of the average highest
backtracking node, $(p_G\simeq 0.78, \alpha _G \simeq 3.0$), coincide with
the analytically computed intersection of the single branch
trajectory and the critical line $\alpha_c(p)$, see Figure~4. As for
complexity, experiments on 3-SAT instances at $\alpha = 3.5$ or on
2+0.78-SAT instances at $\alpha _G =3.0$ are in good
agreement with theoretical calculations of $\omega$, see Table~1. As a
conclusion, the structure of search trees for 3-SAT reflects the
existence of a critical line for 2+p-SAT instances. The latter may be
found back as the locus of highest backtracking nodes reached from all
initial 3-SAT ratios $\alpha$ in the intermediate range.


In conclusion, we have shown that statistical physics tools can be
useful to study the solving complexity of branch and bound algorithms
\cite{NPC,Hay} applied to hard combinatorial optimization or decision
problems.  The phase diagram of Figure~4 affords a qualitative
understanding of the probabilistic complexity of DPLL variants on
random instances. This view may reveal the nature of the complexity of
search algorithms for SAT and related NP-complete problems \cite{NPC}. 
In the sat phase, branch trajectories are related to polynomial time 
computations while in the unsat region, tree trajectories lead to exponential
calculations. Depending on the starting point (ratio $\alpha$ of the
3-SAT instance), one or a mixture of these behaviours is observed.
Figure~4 furthermore gives some insights to improve the search
algorithm. In the unsat region, trajectories must be as horizontal as
possible (to minimize their length) but resolution is necessarily
exponential\cite{Chv}. In the sat domain, heuristics making
trajectories steeper could avoid the critical line $\alpha _C (p)$ and
solve 3-SAT polynomially up to threshold.

\vskip .3cm \noindent
{\bf Acknowledgements:} We are grateful to J. Franco for his 
encouragements and numerous suggestions in the writing of this work. 
We also thank A. Hartmann and M. Weigt for their interest in our results.


\newpage
\vskip 2cm
\begin{center}
{\Large TABLE}
\end{center}
\vskip 1cm

\begin{table}
$$
\begin{array}{|c |c| c|c|}
\hline 
\multicolumn{1}{|c}{\hbox{\rm Ratio} \ \alpha} & 
\multicolumn{2}{|c}{\hbox{\rm Experiments}} & 
\multicolumn{1}{|c|}{\hbox{\rm Theory}} \\
\hbox{\rm clause/var.} & \hbox{\rm nodes} & \hbox{\rm branches} &  \\ 
\hline
20& 0.0153 \pm 0.0002 & 0.0151 \pm 0.0001  & 0.0152  \\
15& 0.0207 \pm 0.0002 & 0.0206 \pm 0.0001  & 0.0206  \\
10& 0.0320 \pm 0.0005 & 0.0317 \pm 0.0002  & 0.0319  \\
7 & 0.0482 \pm 0.0005 & 0.0477 \pm 0.0005  & 0.0477  \\
4.3& 0.089 \pm 0.001  & 0.0895 \pm 0.001   & 0.0875  \\
\hline
3.5& 0.045 \pm 0.005  & 0.044  \pm 0.005   &         \\
\cline{1-1}
3.0     & 0.042 \pm 0.002 & 0.041 \pm 0.003 & 0.0453 \\      
(p=0.78) & & & \\
\hline 
\end{array}
$$
\vskip 1cm
{\bf Table 1:}
Logarithm of the complexity $\omega$ from measures of search tree
sizes (nodes and branches) and theory\cite{nota1}. Experimental
results obtained for $\alpha =3.5$ (sat phase) are compared to the
complexity of the unsat tree built from hit point G
($p_G=0.78$,$\alpha_G=3$), see Figure~4. Uncertainties on $\omega$ are
larger due to finite size critical fluctuations on the location of G.
For $\alpha \gg 1$, our theoretical prediction $\omega = (3+\sqrt{5})
[\ln ((1+\sqrt 5)/2)]^2/ (6\ln 2)/\alpha \simeq 0.292/\alpha$ seems to
be exact \cite{Bea}; indeed, the search tree becomes small enough to
justify the decorrelation of branches done in (\ref{croi}).

\end{table}

\newpage
\vskip 1cm
\begin{center}
{\Large FIGURE CAPTIONS}
\end{center}
\vskip 1cm
{\bf Figure 1:}
Example of 3--SAT instance and Davis--Putnam--Loveland--Logemann
resolution.  {\bf Step~0.}  The instance consists of $M=5$ clauses
involving $N=4$ variables that can be assigned to true (T) or false
(F). $\bar w$ means (NOT $w$) and v denotes the logical OR. The search
tree is empty.  {\bf 1.}  DPLL randomly selects a variable among the
shortest clauses and assigns it to satisfy the clause it belongs to,
e.g. $w=$T (splitting with the Generalized Unit Clause --GUC--
heuristic) [6,9].
A node and an edge symbolising respectively the variable chosen ($w$)
and its value (T) are added to the tree.  {\bf 2.}  The logical
implications of the last choice are extracted: clauses containing $w$
are satisfied and eliminated, clauses including $\bar w$ are
simplified and the remaining ones are left unchanged. If no unitary
clause ({\em i.e.} with a single variable) is present, a new choice of
variable has to be made.  {\bf 3.}  Splitting takes over. Another node
and another edge are added to the tree.  {\bf 4.}  Same as step 2 but
now unitary clauses are present.  The variables they contain have to
be fixed accordingly.  {\bf 5.}  The propagation of the unitary
clauses results in a contradiction. The current branch dies out and
gets marked with C.  {\bf 6.}  DPLL backtracks to the last split
variable ($x$), inverts it (F) and creates a new edge.  {\bf 7.}  Same
as step 4.  {\bf 8.}  The propagation of the unitary clauses
eliminates all the clauses. A solution S is found and the instance is 
satisfiable.  For an
unsatisfiable instance, unsatisfiability is proven when backtracking
(see step 6) is not possible anymore since all split variables have
already been inverted. In this case, all the nodes in the final search
tree have two descendent edges and all branches terminate by a
contradiction C.

\vskip 1cm \noindent
{\bf Figure 2:} 
Types of search trees generated by the DPLL solving procedure. 
{\bf A.} {\em simple branch:} the algorithm finds
easily a solution without ever backtracking. {\bf B.} {\em dense tree:}
in the absence of solution, the algorithm builds a ``bushy'' tree,
with many branches of various lengths, before stopping.  {\bf C.} {\em
mixed case, branch + tree:} if many contradictions arise before
reaching a solution, the resulting search tree can be decomposed in a
single branch followed by a dense tree. The junction G is the highest
backtracking node reached back by DPLL.

\vskip 1cm \noindent
{\bf Figure 3:} 
Complexity of 3-SAT solving for three problem sizes and averaged 
over 10,000 randomly drawn samples. 

\vskip 1cm \noindent
{\bf Figure 4:}
Phase diagram of 2+p-SAT and dynamical trajectories
of DPLL.
The threshold line $\alpha_C (p)$ (bold full line) separates sat
(lower part of the plane) from unsat (upper part) phases. Extremities
lie on the vertical 2-SAT (left) and 3-SAT (right) axis at coordinates
($p=0,\alpha _C=1$) and ($p=1,\alpha _C\simeq 4.3$) respectively.  
Departure points for DPLL trajectories are located on the
3-SAT vertical axis and the corresponding values of $\alpha$ are
explicitly given. Dashed curves represent tree trajectories in the
unsat region (thick lines, black arrows) and branch
trajectories [9] in the sat phase (thin lines, empty
arrows). Arrows indicate the direction of "motion" along trajectories
parametrised by the fraction $t$ of variables set by DPLL.  For small
ratios $\alpha < \alpha _L$, branch trajectories remain
confined in the sat phase and end in S of coordinates $(1,0)$, where a
solution is found. At $\alpha_L\simeq 3.003$, the single branch trajectory hits
tangentially the threshold line in T of coordinates $(2/5,5/3)$. In
the intermediate range $\alpha _L < \alpha < \alpha_C$, the branch
trajectory intersects the threshold line at some point G (that depends
on $\alpha$). A dense tree then grows in the unsat phase, as happens
when 3-SAT departure ratios are above threshold $\alpha > \alpha _C
\simeq 4.3$. The tree trajectory halts on the dot-dashed curve
$\alpha \simeq 1.259/(1-p)$ where the tree growth process stops, see text.
At this point, DPLL has reached back the highest backtracking node in
the serach tree, that is, the first node when $\alpha > \alpha _C$, or node G
for $\alpha _L < \alpha < \alpha_C$.
In the latter case, a solution can be reached from a new descending branch
while, in the former case, unsatisfiability is proven.

\begin{center} 
\begin{figure}
\includegraphics[width=330pt,angle=0]{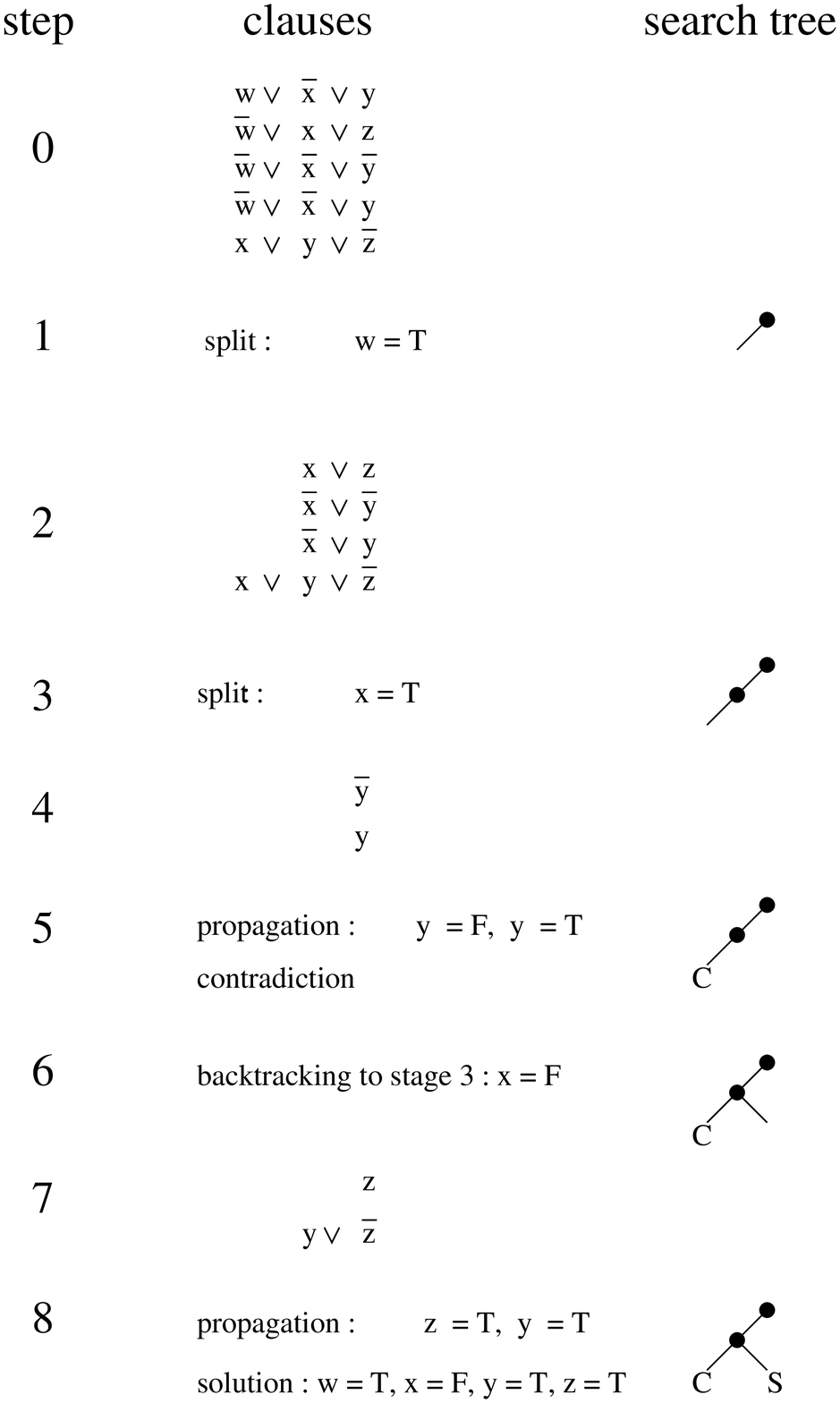}
\vskip 2cm
{\bf Figure 1}
\end{figure}

\newpage
\begin{figure}
\includegraphics[width=250pt,angle=-90]{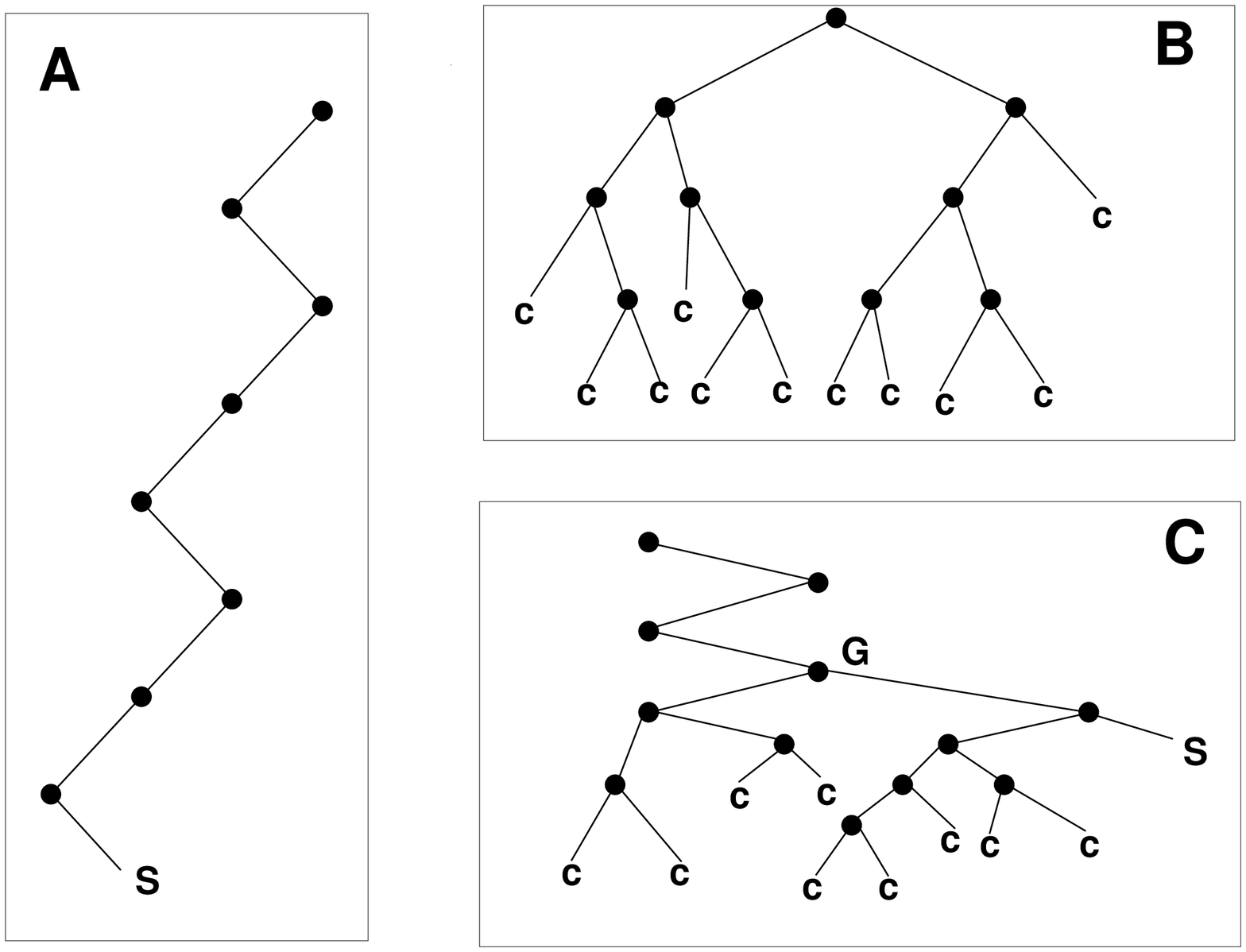}
\vskip 2cm
{\bf Figure 2}
\end{figure}

\newpage
\begin{figure}
\includegraphics[width=280pt,angle=-90]{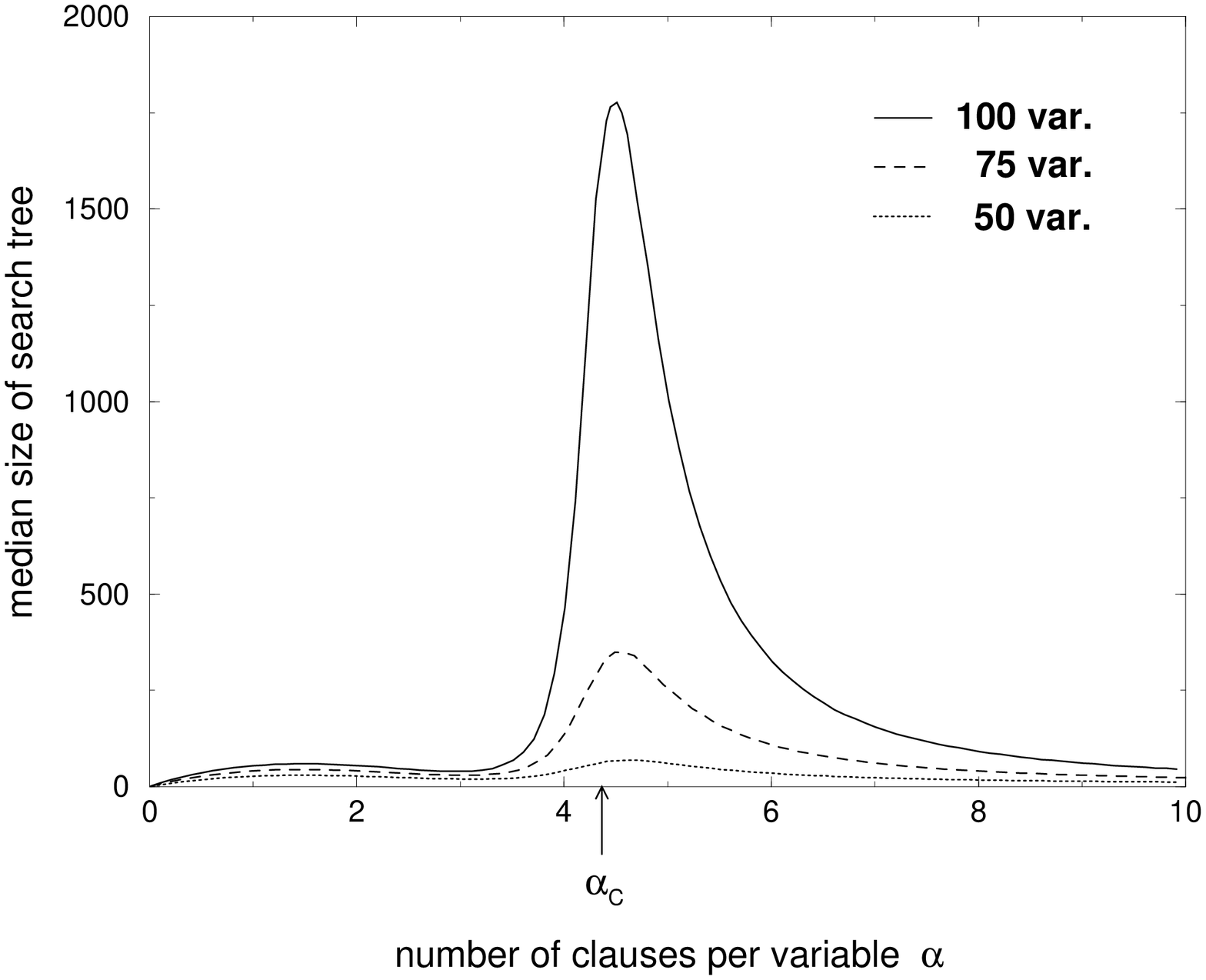}
\vskip 2cm
{\bf Figure 3}
\end{figure}

\newpage
\begin{figure}
\includegraphics[width=330pt,angle=-90]{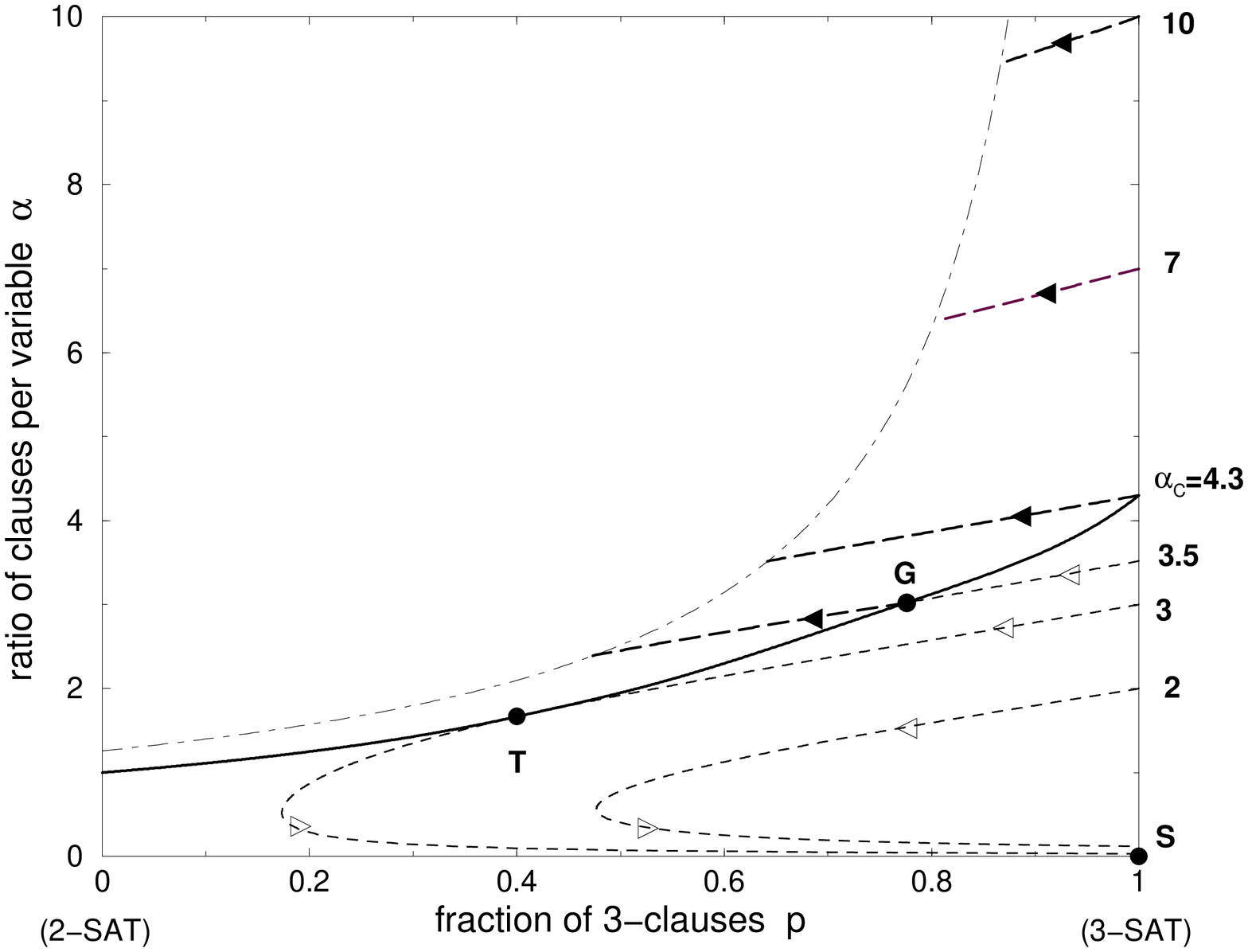}
\vskip 2cm
{\bf Figure 4}
\end{figure}
\end{center}



\begin{thebibliography}{99}
\bibitem{nota1}
In terms of the densities of 2- and 3-clauses, $c_2\,=\,\alpha\,
(1-t)\,(1-p)$, $c_3\,=\,\alpha\ (1-t)\,p\;$, and of the partial
derivatives $z_2=\partial \omega/ \partial c_2$, $z_3=\partial \omega/
\partial c_3$, we have for the GUC heuristics ${\cal H} (c_2 ,c_3, z_2
, z_3 ; t )= \log_{2} \nu (z_2 ) +  [3\; c_3\; ( e^{z_3}
(1+e^{-z_2})/2 -1 )+ c_2 \; (\nu (z_2) -2 ) ] / (1-t) /\log 2$ 
where $\nu (z_2 ) = e^{z_2} \left(1+\sqrt{4\,e^{-z_2}+1}\right)/2$.
We have solved eqn.~(\ref{croi}) by expanding ${\cal H}$ at the linear
order in $z_2,z_3$. Notice that theory concentrates on the 
logarithm of the average number of branches. This annealed 
approximation of $\omega$ is very accurate above threshold \cite{ric}.
\end{thebibliography}
\end{document}